\documentclass{article}
\usepackage{ijcai22}

\usepackage{times}
\usepackage{soul}
\usepackage{url}
\usepackage[hidelinks]{hyperref}
\usepackage[small]{caption}
\usepackage{graphicx}
\usepackage{amsmath}
\usepackage{amsthm}
\usepackage{booktabs}
\usepackage{algorithm}
\usepackage{algorithmic}
\usepackage{multirow}
\usepackage{graphicx}
\usepackage{diagbox}
\usepackage{amssymb}

\usepackage{multirow}
\usepackage{graphicx}

\usepackage{float}                  
\usepackage{subfig}                 
\usepackage{overpic}                
\usepackage[T1]{fontenc} 

\usepackage{color}
\usepackage{url}
\newcommand{\MN}{CLDR}

\newtheorem{theorem}{Theorem}

\usepackage{xcolor}
\usepackage[most]{tcolorbox}

\title{CLDR: Contrastive Learning Drug Response Models from Natural Language Supervision }

\author{
Kun Li$^1$
\and
Wenbin Hu$^1$\
\affiliations
$^1$School of Computer Science, Wuhan University
\emails
\{li\_\_kun, hwb\}@whu.edu.cn
}

\begin{document}

\maketitle

\begin{abstract}

Deep learning-based drug response prediction (DRP) methods can accelerate the drug discovery process and reduce R\&D costs. Although the mainstream methods achieve high accuracy in predicting response regression values,  the regression-aware representations of these methods are fragmented and fail to capture the continuity of the sample order. This phenomenon leads to models optimized to sub-optimal solution spaces, reducing generalization ability and may result in significant wasted costs in the drug discovery phase.  In this paper, we propose \MN, a contrastive learning framework with natural language supervision for the DRP. The \MN~converts regression labels into text, which is merged with the captions text of the drug response as a second modality of the samples compared to the traditional modalities (graph, sequence). In each batch, two modalities of one sample are considered positive pairs and the other pairs are considered negative pairs. At the same time, in order to enhance the continuous representation capability of the numerical text, a common-sense numerical knowledge graph is introduced. We validated several hundred thousand samples from the Genomics of Drug Sensitivity in Cancer dataset, observing the average improvement of the DRP method ranges from 7.8\% to 31.4\% with the application of our framework. The experiments prove that the \MN~effectively constrains the samples to a continuous distribution in the representation space, and achieves impressive prediction performance with only a few epochs of fine-tuning after pre-training. The code is available at: \url{https://gitee.com/xiaoyibang/clipdrug.git}.

\end{abstract}

\section{Introduction}



\begin{figure}
    \centering
    \includegraphics[width=0.99\linewidth]{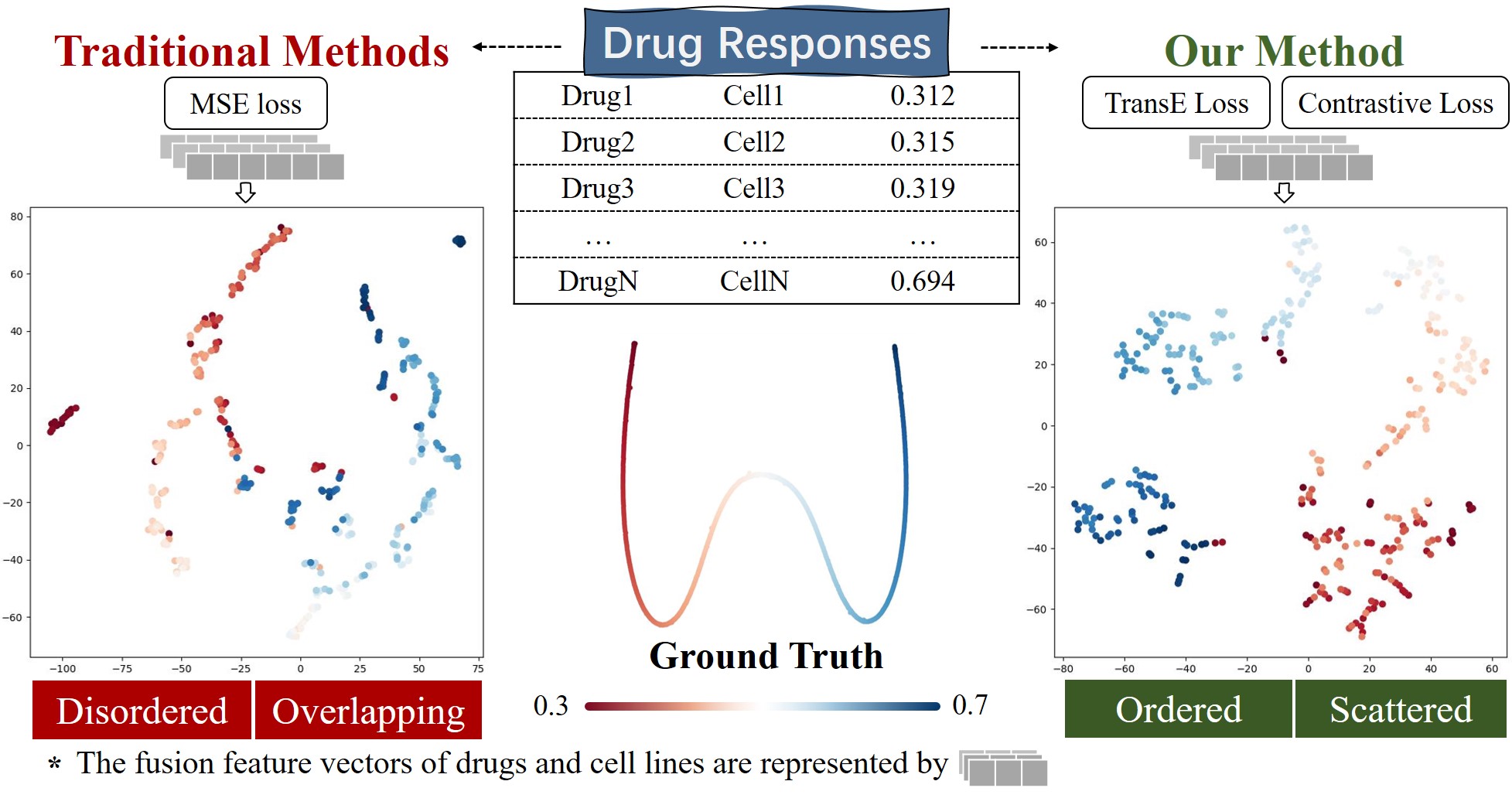}
    \caption{Learned representations of different methods on a real-world drug response regression task. The loss function in general regression learning ($L_1$ loss) fails to recognize the underlying continuous information. In contrast, our method learns continuous representations to capture intrinsic sample order based on the regression task.}
    \label{fig:intro}
\end{figure}

Drug response prediction is the essential step in preclinical screening for Phenotypic Drug Discovery (PDD) \cite{Nature_PDD,Nature_PDD_2}, which is gradually becoming more mature as a field and a recognized discovery paradigm in academia and the pharmaceutical industry, rather than a short-lived craze \cite{PDD_3}. This sustained interest stems from notable successes over the past decade, including SEP-363856 for the treatment of schizophrenia \cite{Drug_1}, KAF156 for malaria \cite{Drug_2}, Crisaborole for atopic dermatitis \cite{Drug_3}, and so on.


In the context of drug response prediction, which is inherently a regression task, the realm of representation learning has traditionally devoted less attention to regression compared to classification. Deep regression models are typically trained end-to-end, lacking explicit emphasis on regression-aware representations. Consequently, the fragmented modeling representation of numerical values and the inability to capture a continuous sequence of samples lead to sub-optimal outcomes across a spectrum of regression tasks. In addition, deep learning-based DRP methods such as \cite{DeepCDR,tcnns,GraTransDRP} can effectively predict the reaction results of drugs with cell lines that have already appeared in the dataset. 

However, in practical applications, the compounds to be tested are unlabelled. Different drugs with different structures and properties have different reaction results in the same cell line \cite{drug_structure}. The prediction performance of DRP under such zero-shot learning conditions decreases drastically, which greatly affects the development of PDD \cite{drug_blind1,drug_blind2}. This situation is mainly due to the poor regression-aware representation of the model, which exhibits disorder, fragmentation, aggregation, and overlapping mappings \cite{ranknc}.

Traditional approaches exhibit shortcomings in representing continuous regression values, particularly under zero-shot conditions, leading to issues of overlap and disorder. These phenomena significantly compromise the performance of DRP models. As shown in Fig.\ref{fig:intro}, the traditional method has a poor representation of continuous regression value samples on the test dataset of Genomics of Drug Sensitivity in Cancer (GDSC) \cite{GDSC}. It can only represent ordered samples that are outside a certain range in an unordered way into the overlapping space. This greatly affects the performance of DRP models under simulations of real-world application scenarios. For example, \cite{110bGCNforDRP} shows that their model GraphDRP prediction performance is as high as 90\% or more when drug species and cell line species are both present in the training set and test set, but the Pearson correlation coefficients (PCC) between the predicted regression values and the labels are only about 4-32\% under the zero-shot condition. Therefore, it becomes an urgent challenge to improve the model's ability to represent continuous values in an ordered and scattered manner, and thus improve the zero-shot generalization performance.

In the natural language processing field, the method of enhancing zero-shot generalization utilizing pre-training has actually been verified to be very successful nowadays, like BERT \cite{BERT} and GPT \cite{GPT1,GPT2}, etc. In the computer vision field, \cite{CLIP} demonstrated that SOTA image representations can be achieved with a simple pre-training task on a dataset of 400 million image-text pairs collected from the Internet.  Therefore, for the drug discovery field, it is possible to construct a link between the drug response data and the labeled text to learn drug response representations from the text, which would then be used for natural language supervised contrast learning to improve the performance of zero-shot learning.


To fill the gap, we present \textbf{\MN~}, a \underline{\textbf{c}}ontrastive learning framework with \underline{\textbf{n}}atural \underline{\textbf{l}}anguage \underline{\textbf{s}}upervision for the DRP. The \MN~converts continuous numerical labels of drug responses into natural language text by prompting templates, and the drug molecules and cell lines are encoded separately for the same sample using natural language methods and traditional methods. The drug and cell line feature space and the labeled text feature space are mapped to a shared representation space through a contrastive learning strategy that maximizes the similarity between related samples and minimizes the similarity between unrelated samples. Then, we construct a common-sense numerical knowledge graph with reference to the definition of ordinal numbers which enhances the perception of continuous values in the representation space of drug and cell line interactions. Through contrastive learning pre-training, ordered and scattered natural language representations are aligned and mapped to a unified high-dimensional space, in conjunction with traditional drug-cell fusion representations. After the pre-training, fine-tuning the model not only yields better performance, but also significantly improves the robustness and generalization ability. It is worth noting that our framework \MN~is orthogonal to existing DRP methods, allowing the use of any regression method that maps the learned representation to the final predicted value.

We provide rigorous theoretical proof of the soundness of the \MN~method in Section \ref{section3}. To support practical evaluations, We validate more than 150,000 samples from the GDSC2, a classical drug response dataset.  The results show that the average improvement of the DRP method ranges from 7.8\% to 31.4\% with the application of our method. The \MN~effectively constructs links between drug response data and labeled text, which in turn enhances the out-of-distribution generalization ability and improves the success rate of phenotypic drug discovery. Our contributions are as follows:



\begin{itemize}

\item We propose the \MN~model, a novel contrastive learning framework with natural language supervision for the DRP task. The \MN~model effectively constructs links between drug response data and labeled text, enhancing out-of-distribution generalization ability and improving the success rate of phenotypic drug discovery.

\item The common-sense numerical knowledge graph is constructed based on ordinal numbers, enhancing the perception of continuous values in the representation space of drug and cell line interactions.

\item The practical evaluations on GDSC2 showcase that the application of the \MN~leads to a notable improvement in DRP methods, enhancing results by at least 7.8\% and up to 31.4\%, highlighting its effectiveness in enhancing generalization and success rates in phenotypic drug discovery.


\end{itemize}

\begin{figure*}[htpb]
    \centering
    \includegraphics[width=1\linewidth]{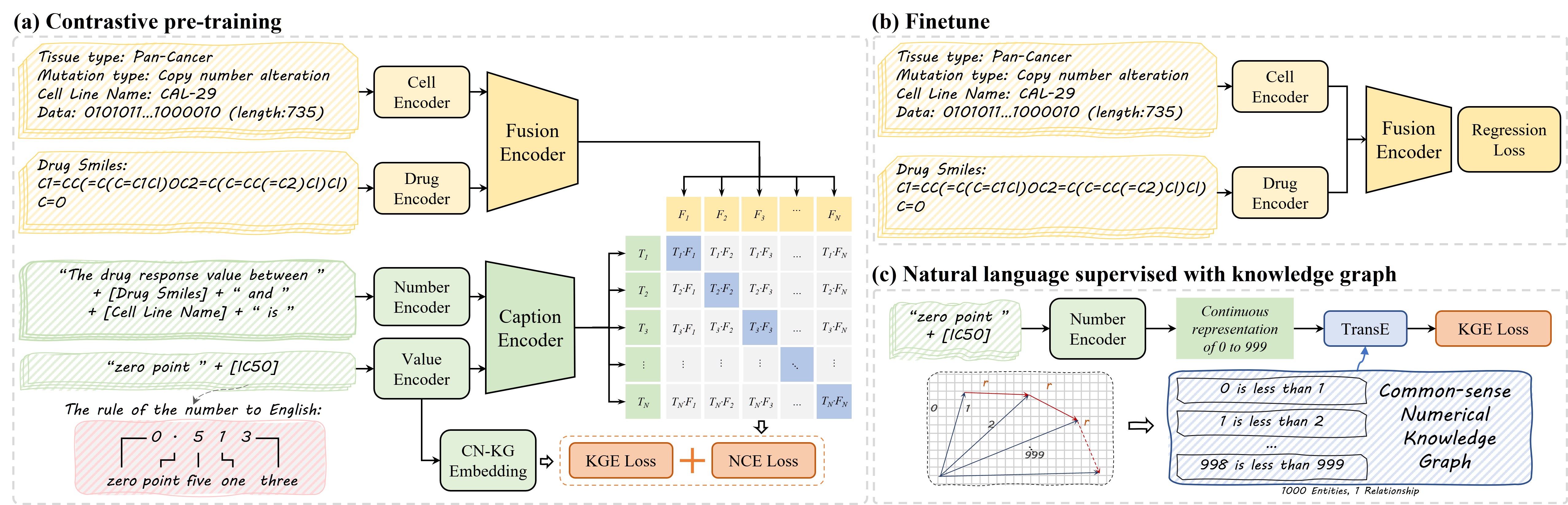}
    \caption{Summary of our method. (a)While a standard drug response prediction model jointly trains a drug-cell feature fusion extractor and a regressor to predict IC50, the \MN~jointly trains the cell and drug fusion encoder and the caption encoder to predict the correct pairing of a batch of (drug response inputs, textual description of the response process) training samples. (b)When fine-tuning, the regressor of the model is trained with the regression loss function. (c)When the caption encoder describes the response process, the number encoder is constrained to learn the features of numerical continuity with the common-sense knowledge graph.}
    \label{fig:framework}
\end{figure*}

\section{Related work}



With the culmination of extensive projects investigating drug response in cell lines, exemplified by GDSC \cite{GDSC} and CCLE \cite{CCLE}, the exploration of DRP has become feasible. Various DRP techniques have been introduced, and categorized into 1DCNN-based, graph-based, and transformer-based methodologies.
Non-graph-based methodologies typically employ Convolutional Neural Networks (CNNs) and Multi-Layer Perceptrons (MLPs) for information retrieval. These methods encode drug molecules in Simplified Molecular Input Line Entry Specification (SMILES) string format and directly extract gene sequence features through 1DCNN, overlooking the pharmacological and structural attributes of the molecule. Conversely, graph-based methods \cite{DeepCDR,110bGCNforDRP} concentrate on converting drug molecules into graph structures, utilizing Graph Neural Networks (GNNs) for graph representation rather than directly extracting features from SMILES strings. Additionally, in contrast to GNNs, the transformer-based methods \cite{GraTransDRP,DeepTTC} avoid introducing any structural inductive bias at intermediate layers \cite{GraTransDRP},  thereby mitigating the expressivity limitations of GNNs \cite{AGCNforDD}. 




\section{Method}

\subsection{Basic Definitions}


Refer to \cite{MSDA} for basic definitions related to the DRP task, the input data are drugs and cell lines, and the output is the half maximal inhibitory concentration ($IC_{50}$). A single drug domain $j$ is defined as a set of response data for a specific drug, $d_j$, across multiple cell lines, denoted as ${c_1^j, c_2^j, \ldots, c_n^j}$, comprising a total of $n$ samples. The drug response data are divided into drug data domains based on drug types, denoted as $S:=\left \{ (X_{sj}, Y_{sj}) \right \}_{j=1}^{N}$; here, $S$ indicates that all drug response data are divided according to $N$ types of drugs. $X_{sj}:=\left \{ (d_{j}^{s},c_{i}^{sj}) \right \}_{i=1}^{\left | {X_{sj}} \right |} $ represents samples from drug domain $j$, while $Y_{sj}:= \left \{ y_{i}^{sj} \right \}_{i=1}^{\left | {X_{sj}} \right |} $ is the corresponding $IC_{50}$ of samples. This zero-shot learning task presents a greater level of complexity compared to the random splitting of the entire dataset commonly employed in supervised learning \cite{NMIDTI}.

\subsection{Fusion Encoder}
This paper does not provide a detailed exposition of the computational methods employed in these two data representation branches. Broadly, we categorize data feature extraction techniques into CNNs, MLPs, GNNs, Transformers, and Transformer-based GNNs.

The input types for the drug extractor are the SMILES sequence or molecular graph. We uniformly describe the process of branching feature representations of the drug $d_i$ and the cell line $c_i$ as follows:

\begin{equation}
\mathbb{P}_i= \Phi_{\mathrm{drug}}(d_i), \mathbb{Q}_i= \Phi_{\mathrm{cell}}(c_i)
\end{equation}
\noindent where, $\Phi_{\mathrm{drug}}$ and $\Phi_{\mathrm{cell}}$  denotes the drug and cell line extractor, while $\mathbb{P}_i$ and $\mathbb{Q}_i$ are the representations of the drug $d_i$ and the cell line $c_i$. The fusion branch can be expressed as follows:

\begin{equation}
    F = \Phi_{\mathrm{f}}( [ \mathbb{P}_i, \mathbb{Q}_i ])
\end{equation}

\noindent where $\Phi_{\mathrm{f}}$ denotes the fusion branch that can be replaced by different methods. The fusion branch of the multi-source domains uses pre-trained model parameters.

\subsection{Caption Encoder}



In the caption encoder, the text-based description is the second coding modality for the drug response process. The inputs $\mathbf{a}$, $\mathbf{b}$ and the output $\mathbf{c}$ in the DRP task are described uniformly with the prompt template $\mathcal{F}_{\mathrm{prompt}}(x)$ as follows:


\textbf{\textit{\textcolor{blue}{The drug response value between   [$\mathbf{a}$]     and   [$\mathbf{b}$]   is } \textcolor{red}{[$\mathbf{c}$]}}}


\noindent where $\mathbf{a}$ denotes the SMILES string representation of the drug, $\mathbf{b}$ denotes the name of the cell line, $\mathbf{c}$ denotes the quantitative output value of the reaction level between the drug and the cell line, i.e., the $IC_{50}$. The $IC_{50}$ will be normalized to between 0 and 1 during pre-processing. 

To represent $IC_{50}$ efficiently, we define a transformation method $\mathcal{F}_{\mathrm{num2str}}(x)$ that converts it to English words in character order, where the decimal point is encoded as 'point'. For example, as shown in Fig.\ref{fig:framework}(a), '0.513' can be converted to "zero point five one three". To facilitate continuity constraints on the values, the caption encoder is designed to consist of two encoders $\Phi_{\mathrm{text}}$, $\Phi_{\mathrm{number}}$, which encode text-based descriptions $T_{\mathrm{text}}=\mathcal{F}_{\mathrm{prompt}}(\mathbf{a}, \mathbf{b})$ and value-based descriptions $T_{\mathrm{number}}=\mathcal{F}_{\mathrm{num2str}}( \mathbf{c} )$, respectively. Then the two feature vectors are merged to fuse the descriptive information of the drug reaction process using the multi-layer Transformer $\Phi_{\mathrm{cap}}$, as follows:

\begin{equation}
T=\Phi_{\mathrm{cap}}\left ( \left [ \Phi_{\mathrm{text}}\left ( T_{\mathrm{text}} \right ) ,\Phi_{\mathrm{number}}\left ( T_{\mathrm{number}} \right ) \right ]   \right ) 
\end{equation}

\noindent where $T\in \mathbb{R}^{N \times m} $ is the description text feature vector after encoding by caption encoder.

\subsection{Natural language supervised with knowledge graph}


To ensure the ability to perceive the continuity of numbers when encoding numerical values in natural language, we construct a common-sense numerical knowledge graph (\textbf{CN-KG}) with reference to \cite{NEKG}, which uses linear structures to construct graphs that correctly and intuitively represent numbers and their magnitude relationships. As shown in Fig.\ref{fig:framework}(c), the entities of the CN-KG are sequences of ordered numbers. Entities in the CN-KG are integers that include a valid decimal part of $\mathcal{C}$, where the minimum and maximum numbers are customized to the specific task. The entities of the graph are linked by a single relationship type called "is less than", which ensures that the transfer properties of the numbers are captured.

When dealing with numerical information, we must consider how to incorporate numerical features into this framework in order to represent numerical relationships more accurately. To enhance the ability of numerical encoders with CN-KG to efficiently capture numerical relational features, we propose a margin-based loss function $\mathcal{L}_{\mathrm{KGE} }$ for the embedding of the CN-KG. The goal is to minimize the difference of embedding vectors between the set of entities $E$ and the set of the relationships $L$. 



\begin{equation}
\mathcal{L}_{\mathrm{KGE}}= \sum_{ (h,l,t)\in S   } \left [ \gamma +d(\textbf{h}+\textbf{l},\textbf{t}) -d(\textbf{t}+\textbf{l} ,\textbf{h})\right ]_+
\end{equation}

\noindent where, $[x]_+$ denotes the positive part of $x$, $\gamma > 0$ is a margin hyperparameter. The set $S$ is composed of the triplets $(h,l,t)$, with $h,t \in E$, $l \in L$. The embeddings $\mathbf{h,l,t}$ take values in $\mathbb{R}^{k}$ ($k$ is a hyperparameter) and are denoted with the same letters in boldface characters. Considering also the specificity of numerical relations, it is necessary to encode the CN-KG entities $h,t$ by the value encoder $\Phi_{\mathrm{number}}$. For the similarity measure $d$, either the $L_1$ or $L_2$-norm can be used.


\noindent \textbf{Pre-training}. For the $i$-th normally represented sample $(d_i,c_i)$ and the $j$-th one described by the language $(d_j,c_j,y_j)$ in a batch $\mathcal{B}$, we normalize their feature vectors in a hyper-sphere using $u _i :=\frac{ \Phi_{\mathrm{f}}(d _i,c _i)}{\left \|  \Phi_{\mathrm{f}}(d _i,c _i) \right \| } $ and $v _j :=\frac{ \Phi_{\mathrm{cap}}(d _j,c _j,y_j)}{\left \|  \Phi_{\mathrm{cap}}(d _j,c _j,y_j) \right \| } $. The similarity between $u _i$ and $v _j $ is calculated as $u_i^\mathrm{T} v _j$.  A supervised contrastive loss function is considered to train the model:

\begin{equation}
    \begin{aligned}
        \mathcal{L}_{\mathrm{NCE}} = -\frac{1}{N} \left ( \sum_{i}^{N}\mathrm{log} \frac{\mathrm{exp} ( u_i^\mathrm{T} v _i/\sigma )}{  {\textstyle \sum_{j=1}^{N}} \mathrm{exp} ( u_i^\mathrm{T} v _j/\sigma )} +  
         \right.
        \\
        \phantom{=\;\;}
        \left. \sum_{i}^{N}\mathrm{log} \frac{\mathrm{exp} ( v_i^\mathrm{T} u_i/\sigma )}{  {\textstyle \sum_{j=1}^{N}} \mathrm{exp} (v_i^\mathrm{T} u_j /\sigma )} \right ) 
    \end{aligned}
\end{equation}

\noindent where, $N$ is the size of the batch $\mathcal{B}$, and $\sigma$ is the temperature to scale the logits.

The goal of the \MN~in the pre-training phase is to jointly optimize the following contrast loss function and the loss function of the CN-KG embedding:

\begin{equation}
\mathcal{L}_{\mathrm{All}}=\alpha \mathcal{L}_{\mathrm{NCE}}+(1-\alpha )\mathcal{L}_{\mathrm{KGE}}
\end{equation}

\noindent where $\alpha$ represents the weight adjustment factor of joint optimization of two loss functions.

\noindent \textbf{Fine-tuning}.




In the fine-tuning stage, we employ the standard Mean Squared Error (MSE) loss function for supervised regression on the normal fusion branch. A regression output layer $\Phi_{\mathrm{mlp}}$ based on the MLP is designed after the fusion branch as follows:
\begin{equation}
    \mathcal{L}_{\mathrm{REG}} = \frac{1}{\left | \mathcal{B} \right |}  {   \sum_{i=0}^{\left | \mathcal{B} \right |  } (  \Phi_{\mathrm{mlp}}(\Phi_{\mathrm{fusion}}(d_i,c_i)) - y_i )^2} 
\end{equation}

\noindent where, $\mathcal{B}$ represents a batch of samples, $y_i$ is the  normalized value of $IC_{50}$ corresponding to $\left \{ d_i, c_i \right \} $.

\section{Theoretical Analysis}
\label{section3}

\begin{table*}[htpb]
\centering
\caption{The overall experiment of our method. Each method shows the results of the original and those based on our method (+\MN), where Improv. denotes the percentage of enhancement.}

\renewcommand \arraystretch{1.2}
\setlength{\tabcolsep}{2mm}{
\resizebox{\textwidth}{!}{%
\begin{tabular}{cccccccccccc}
\hline
\multicolumn{2}{c}{} & \multicolumn{5}{c}{Drug} & \multicolumn{5}{c}{Total} \\
\multicolumn{2}{c}{\multirow{-2}{*}{Methods}} & RMSE $\downarrow$ & MSE $\downarrow$ & PCC $\uparrow$ & SPC $\uparrow$ & Rank $\downarrow$ & RMSE $\downarrow$ & MSE $\downarrow$ & PCC $\uparrow$ & SPC $\uparrow$ & Rank $\downarrow$ \\ \hline
 \multirow{3}{*}{\begin{tabular}[c]{@{}c@{}}tCNNs\\ \cite{tcnns} \end{tabular}} & Original & 0.0548 & 0.0036 & 0.4710 & 0.4682 & 0.0257 & 0.0596 & 0.0036 & 0.5342 & 0.4632 & 0.0267 \\
 & +\MN~& \textbf{0.0539} & \textbf{0.0034} & \textbf{0.5272} & \textbf{0.5321} & \textbf{0.0218} & \textbf{0.0580} & \textbf{0.0034} & \textbf{0.5451} & \textbf{0.4894} & \textbf{0.0227} \\
 & (Improv.) & 1.64\% & 5.56\% & 11.93\% & 13.65\% & 15.18\% & 2.71\% & 5.34\% & 2.05\% & 5.64\% & 15.13\%\\ \hline
 & Original & 0.0620 & 0.0057 & 0.4405 & 0.4409 & 0.0291 & 0.0729 & 0.0054 & 0.3231 & 0.1986 & 0.0302 \\
 & +\MN~& \textbf{0.0590} & \textbf{0.0044} & \textbf{0.4778} & \textbf{0.4797} & \textbf{0.0200} & \textbf{0.0659} & \textbf{0.0043} & \textbf{0.3873} & \textbf{0.2756} & \textbf{0.0208} \\
\multirow{-3}{*}{\begin{tabular}[c]{@{}c@{}}DeepTTC\\ \cite{DeepTTC}\end{tabular}} & (Improv.) & {\color[HTML]{000000} 4.84\%} & {\color[HTML]{000000} 22.81\%} & {\color[HTML]{000000} 8.47\%} & {\color[HTML]{000000} 8.80\%} & {\color[HTML]{000000} 31.27\%} & {\color[HTML]{000000} 9.55\%} & {\color[HTML]{000000} 19.16\%} & {\color[HTML]{000000} 19.89\%} & {\color[HTML]{000000} 38.75\%} & {\color[HTML]{000000} 31.32\%} \\ \hline
 & Original & 0.0598 & 0.0050 & 0.4599 & 0.4546 & 0.0330 & 0.0726 & 0.0053 & 0.3419 & 0.2282 & 0.0342 \\
 & +\MN~& \textbf{0.0560} & \textbf{0.0040} & \textbf{0.5360} & \textbf{0.5402} & \textbf{0.0264} & \textbf{0.0640} & \textbf{0.0041} & \textbf{0.4262} & \textbf{0.2696} & \textbf{0.0275} \\
\multirow{-3}{*}{\begin{tabular}[c]{@{}c@{}}DeepCDR\\ \cite{DeepCDR}\end{tabular}} & (Improv.) & {\color[HTML]{000000} 6.35\%} & {\color[HTML]{000000} 20.00\%} & {\color[HTML]{000000} 16.55\%} & {\color[HTML]{000000} 18.83\%} & {\color[HTML]{000000} 20.00\%} & {\color[HTML]{000000} 11.82\%} & {\color[HTML]{000000} 22.25\%} & {\color[HTML]{000000} 24.68\%} & {\color[HTML]{000000} 18.13\%} & {\color[HTML]{000000} 19.66\%} \\ \hline
 & Original & 0.0637 & 0.0056 & 0.4402 & 0.4447 & 0.0337 & 0.0729 & 0.0053 & 0.3096 & 0.2037 & 0.0312 \\
 & +\MN~& \textbf{0.0565} & \textbf{0.0047} & \textbf{0.5285} & \textbf{0.5348} & \textbf{0.0287} & \textbf{0.0699} & \textbf{0.0049} & \textbf{0.3410} & \textbf{0.2712} & \textbf{0.0297} \\
\multirow{-3}{*}{\begin{tabular}[c]{@{}c@{}}GraphDRP\\ \cite{GraphDRP}\end{tabular}}  & (Improv.) & {\color[HTML]{000000} 11.38\%} & {\color[HTML]{000000} 16.23\%} & {\color[HTML]{000000} 20.05\%} & {\color[HTML]{000000} 20.25\%} & {\color[HTML]{000000} 14.91\%} & {\color[HTML]{000000} 4.09\%} & {\color[HTML]{000000} 7.77\%} & {\color[HTML]{000000} 10.14\%} & {\color[HTML]{000000} 33.14\%} & {\color[HTML]{000000} 4.91\%} \\ \hline
 & Original & 0.0593 & 0.0047 & 0.4738 & 0.4770 & 0.0283 & 0.0666 & 0.0044 & 0.4080 & 0.3255 & 0.0292 \\
 & +\MN~& \textbf{0.0523} & \textbf{0.0039} & \textbf{0.5288} & \textbf{0.5333} & \textbf{0.0264} & \textbf{0.0633} & \textbf{0.0040} & \textbf{0.4665} & \textbf{0.4424} & \textbf{0.0274} \\
\multirow{-3}{*}{\begin{tabular}[c]{@{}c@{}}GratransDRP\\ \cite{GraTransDRP}\end{tabular}}  & (Improv.) & {\color[HTML]{000000} 11.80\%} & {\color[HTML]{000000} 17.02\%} & {\color[HTML]{000000} 11.61\%} & {\color[HTML]{000000} 11.80\%} & {\color[HTML]{000000} 6.71\%} & {\color[HTML]{000000} 5.01\%} & {\color[HTML]{000000} 9.77\%} & {\color[HTML]{000000} 14.33\%} & {\color[HTML]{000000} 35.92\%} & {\color[HTML]{000000} 6.39\%} \\ \hline
 & Original & 0.0624 & 0.0052 & 0.5060 & 0.5040 & 0.0295 & 0.0694 & 0.0048 & 0.3040 & 0.1635 & 0.0331 \\
 & +\MN~& \textbf{0.0547} & \textbf{0.0038} & \textbf{0.5149} & \textbf{0.5294} & \textbf{0.0229} & \textbf{0.0612} & \textbf{0.0037} & \textbf{0.4768} & \textbf{0.3721} & \textbf{0.0239} \\
\multirow{-3}{*}{\begin{tabular}[c]{@{}c@{}}TransEDRP\\ \cite{TransEDRP}\end{tabular}}  & (Improv.) & 12.38\% & 26.23\% & 1.77\% & 5.05\% & 22.42\% & 11.93\% & 22.43\% & 56.86\% & 127.62\% & 27.92\% \\ \hline
\end{tabular}%
}
}
\label{tab:overall}
\end{table*}

In this section, we theoretically prove that joint optimization of $\mathcal{L}_{\mathrm{NCE}}$ and $\mathcal{L}_{\mathrm{KGE}}$  enables fusion branches $\Phi_{\mathrm{f}}$ to obtain continuous regression-aware representation.  All proofs will be in the Appendix.

\noindent \textbf{Notations}. Let $x_i,y_i$ from one drug domain $j$ be noted as inputs and outputs respectively, where $y_i$ is a sorted label with $y_i<y_{i+1}$ constraints, and $x_i \equiv x_{i+1}$. The $ \delta \in (0,1) $ denotes the minimum interval of normalized labels $y_i\in[0,1]$, i.e., the upper bound on the accuracy of the predicted values, then we have $\mathrm{ min} (y_i - y_{i+1}) \ge \delta$. 


First, based on the loss function $\mathcal{L}_{\mathrm{KGE}}$ and the expectation $h+r\approx t$,i.e., $d(\Phi_{\mathrm{number}}(y_i),\Phi_{\mathrm{number}}(y_{i+1})) :=  \mathbf{l}  +\epsilon $. $\mathbf{l}$ is the only kind of relationship embedding in the CN-KG and $\epsilon \in \mathbb{R}$ is one perturbation of the model. We denote $\Phi_{\mathrm{number}}$ as $\mathcal{N}$, then the following theorem can be inferred:



\begin{tcolorbox}[colback=gray!10,colframe=white]
    \label{theorem1}
    \textcolor{black}
    {
    \begin{theorem}[Equal interval representation of $\mathcal{N}$]
       For any $0< \delta <1$,  there exists two perturbations $\epsilon_0,\epsilon_1$ that make the differences of continuous numerical representation the same, 
       \begin{center}$ \mathcal{N}(y_i)-\mathcal{N}(y_{i+1})  = \epsilon_0 \cdot C+ \epsilon_1$.
       \end{center}
    \end{theorem}
    }
\end{tcolorbox}

Next, based on the $\mathcal{L}_{\mathrm{NCE}}$ constraints on positive $x_i, y_i$ and negative samples $x_i^-, y_i^-$, we can obtain the expectation:



\begin{equation}
    \begin{aligned}
        d(\mathcal{C} (x_i, y_i), \mathcal{F}({x_i}) ) &\ll d(\mathcal{C} (x_i^-, y_i^-), \mathcal{F}({x_i}) )
        \\
        d(\mathcal{C} (x_i, y_i), \mathcal{F}({x_i}) ) &\ll d(\mathcal{C} (x_i, y_i), \mathcal{F}({x_i^-}) ) 
    \end{aligned}
\end{equation}

\noindent where $\mathcal{C}$ and $\mathcal{F}$ denote $\Phi_{\mathrm{cap}}$ and $\Phi_{\mathrm{f}}$, respectively.  let $\mathcal{M}:= \mathcal{C} (x_i, y_i)- \mathcal{F}({x_i}) $, we have:

\begin{tcolorbox}[colback=gray!10,colframe=white]
    \label{theorem2}
    \textcolor{black}
    {
    \begin{theorem}[Lower bound of $\mathcal{L}_{\mathrm{NCE}}$]
       For any $i\in j$,  there exist  $\epsilon_2 >0$ such that 
       $\left \| \mathcal{M} \right \| \le \epsilon _2$.
    \end{theorem}
    }
\end{tcolorbox}

\begin{tcolorbox}[colback=gray!10,colframe=white]
    \textcolor{black}
    {
    \begin{theorem}[Main theorem]
        For any $i\in j$,  there exist  $\theta \in (0,1)$ such that if $y_i\to y_{i+1}$, then 
        \begin{center}
              $\left \| \mathcal{F}(x_i) - \mathcal{F}(x_{i+1}) \right \|  \le  2\epsilon _2$.
        \end{center}
    \end{theorem}
    }
\end{tcolorbox}

More generally, let  $\mathcal{Q} := \mathcal{F}(x_i) - \mathcal{F}(x_{i+1})$, then we have the following expression:
\begin{equation}
    \label{equation_end}
    \begin{aligned}
        \left \| \mathcal{Q} \right \|  \le  \left \| \frac{\partial \mathcal{C}}{\partial  y} \left ( x_i,\theta  y_i+  \left ( 1-\theta \right ) y_{i+1} \right )   \right \|     \left ( y_{i+1}-y_i \right ) +2\epsilon _2
    \end{aligned}
\end{equation}

As a result, we prove that joint optimization of $\mathcal{L}_{\mathrm{NCE}}$ and $\mathcal{L}_{\mathrm{KGE}}$  enables fusion branches $\Phi_{\mathrm{f}}$ to obtain continuous regression-aware representation. In addition, it can be observed that lowering the $\delta$ and $\mathcal{L}_{\mathrm{NCE}}$ can improve the ability of continuous regression-aware representation of $\Phi_{\mathrm{f}}$.

\section{Experiments}

\subsection{Experiment Settings}



\noindent \textbf{Validation strategy}. In the context of zero-shot learning experiments, to simulate the practical application scenario of preclinical drug screening, the response data is clustered by drug type and then randomly partitioned into source and target domains, with the type of drug serving as the criterion for division. This zero-shot learning task presents a greater level of complexity compared to the random splitting of the entire dataset commonly employed in supervised learning \cite{NMIDTI}. 

\noindent \textbf{Metrics}. To comprehensively evaluate the impact of the \MN, we employed several evaluation metrics: Root Mean Square Error (RMSE) to gauge deviation, Pearson correlation coefficient (PCC) for assessing linear correlation, Spearman's Rank Correlation Coefficient (SRC) to measure monotonicity, and Margin Ranking Loss (Rank) to evaluate ranking performance.

\subsection{Overall experiment}


This paper proposes the \MN~which aims to leverage the powerful representation of text and the numerical commonsense knowledge graph of ordered numbers to align drug response feature extraction networks, thus improving the generalization ability of the DRP model for zero-sample learning. To this end, we select DRP methods based on deep learning models such as CNNs, GNNs, Transformers, etc., to verify the effectiveness and generality of the \MN. In the overall experiment, we select six methods and test them on the most widely employed GDSC2 dataset. As shown in Table \ref{tab:overall}, the prediction results of the original models are denoted by \textbf{Original}, while results after integrating the \MN~are labeled \textbf{+\MN~}. All methods show increases of 7.88\%, 19.49\%, 17.83\%, 14.29\%, 13.04\%, and 31.46\% after adopting our framework the \MN. The results show that the \MN~can be generally applied to various DRP methods to enhance method performance in zero-shot learning.


\subsection{Ablation study}

\begin{table*}[htpb]
\centering
\caption{The ablation experiments of our method. The design rationality of our method is tested on the GDSC dataset by employing TransEDRP as a drug-response fusion encoder. All numerical values in the table are rounded to three significant figures.}
\label{tab:ablation}
\renewcommand \arraystretch{1.2}
\setlength{\tabcolsep}{1mm}{
\resizebox{\textwidth}{!}{%
\begin{tabular}{cccccccccccccccc}
 \hline
 
\multirow{2}{*}{Finetune} & \multirow{2}{*}{Caption} & \multirow{2}{*}{Precision} & \multicolumn{3}{c}{Loss} & \multicolumn{5}{c}{Total} & \multicolumn{5}{c}{Drug} \\ \cline{4-6}
 &  &  & MSE & CNE & KGE &  RMSE $\downarrow$ & MSE $\downarrow$ & PCC $\uparrow$ & SPC $\uparrow$ & Rank $\downarrow$ & RMSE $\downarrow$ & MSE $\downarrow$ & PCC $\uparrow$ & SPC $\uparrow$ & Rank $\downarrow$  \\ \hline
× & - & - & $\surd$ &  &  & 0.069 & 0.005 & 0.304 & 0.163 & 0.033 & 0.061 & 0.005 & 0.494 & 0.494 & 0.032 \\ \hline
× & Number & 0.001 &  & $\surd$ &  & 0.089 & 0.008 & 0.299 & 0.248 & 0.038 & 0.077 & 0.008 & 0.434 & 0.466 & 0.037 \\
$\surd$ & Number & 0.001 & $\surd$ &  &  & 0.063 & 0.004 & 0.414 & 0.361 & 0.025 & 0.056 & 0.004 & 0.516 & 0.525 & 0.024 \\ \hline
× & Number & 0.001 &  & $\surd$ & $\surd$ & 0.084 & 0.007 & 0.262 & 0.150 & 0.041 & 0.073 & 0.007 & 0.446 & 0.467 & 0.040 \\
$\surd$ & Number & 0.001 & $\surd$ &  &  & 0.065 & 0.004 & 0.403 & 0.301 & 0.027 & 0.055 & 0.004 & 0.514 & 0.520 & 0.026 \\ \hline
× & Text+Number & 0.1 &  & $\surd$ & $\surd$ & 0.451 & 0.203 & 0.006 & 0.051 & 0.156 & 0.132 & 0.054 & 0.008 & 0.006 & 0.048 \\
× & Text+Number & 0.01 &  & $\surd$ & $\surd$ & 0.428 & 0.183 & -0.085 & -0.175 & 0.143 & 0.197 & 0.084 & 0.008 & 0.004 & 0.040 \\
× & Text+Number & 0.001 &  & $\surd$ & $\surd$ & 0.332 & 0.110 & 0.167 & 0.178 & 0.078 & 0.238 & 0.081 & -0.011 & -0.009 & 0.076 \\ 
× & Text+Number & 0.001 &  & $\surd$ &  & 0.311 & 0.097 & -0.279 & -0.211 & 0.112 & 0.243 & 0.085 & -0.008 & -0.005 & 0.096 \\ \hline
$\surd$ & Text+Number & 0.1 & $\surd$ & $\surd$ & $\surd$ & 0.065 & 0.004 & 0.411 & 0.287 & 0.023 & 0.056 & 0.004 & 0.519 & 0.527 & 0.022 \\
$\surd$ & Text+Number & 0.01 & $\surd$ & $\surd$ & $\surd$ & 0.061 & 0.004 & 0.484 & 0.362 & 0.024 & 0.054 & 0.004 & 0.531 & 0.540 & 0.023 \\ 
$\surd$ & Text+Number & 0.001 & $\surd$ & $\surd$ & $\surd$ & 0.061 & 0.004 & 0.502 & 0.354 & 0.024 & 0.056 & 0.004 & 0.517 & 0.523 & 0.023 \\ \hline
$\surd$ & Text+Number & 0.001 & $\surd$ & $\surd$ &  & 0.066 & 0.004 & 0.371 & 0.321 & 0.027 & 0.057 & 0.004 & 0.521 & 0.528 & 0.026 \\ \hline

\end{tabular}%
}
}
\end{table*}

\begin{figure*}[ht!]
    \centering
    \includegraphics[width=0.9\textwidth]{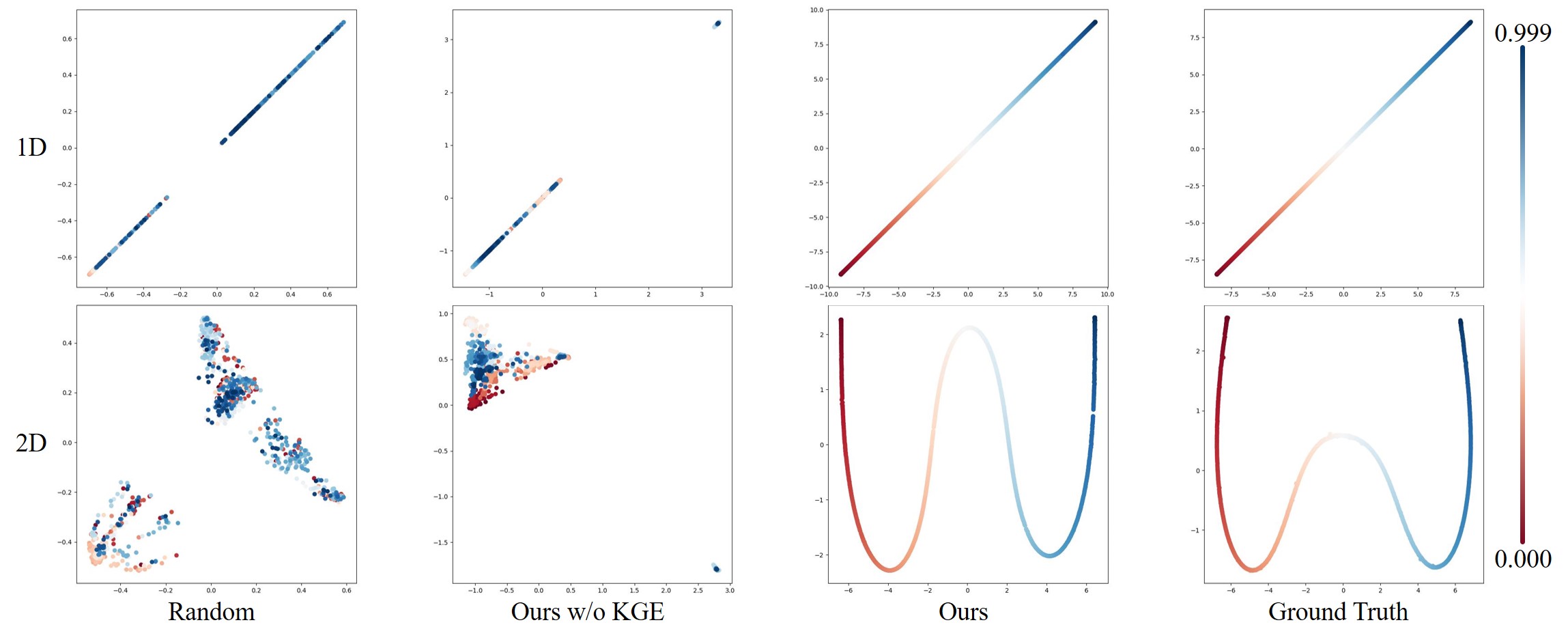}
    \caption{These figures display the learned representations of 0.000 to 0.999 with different strategies, where the text representations are reduced in 1D and 2D by t-distributed stochastic neighbor embedding for comparing the effect of CN-KG constraints.}
    \label{fig:ablation}
\end{figure*}

\noindent \textbf{\textbf{CN-KG}}. To verify that the CN-KG can enhance the continuous representation capability of the numerical text, we design the experiment without exploiting the knowledge graph embedding. Specifically, we do not constrain the number encoder with $\mathcal{L}_{\mathrm{KGE}}$ and keep the pre-training and then fine-tuning approach unchanged. Based on the ablation experiments in Table \ref{tab:ablation}, it can be calculated that the model's performance increases by 9.6\% when incorporating the $\mathcal{L}_{\mathrm{KGE}}$. Furthermore, as shown in Fig.\ref{fig:ablation}, the vectors of numerical text representations whether employing the knowledge graph embedding are compared, since the contrastive learning enables the fusion branch to learn the continuous numerical representation capability of the number encoder.  Although the fusion branch without the CN-KG constraint can still represent the continuous information of the samples in a more scattered manner, it is unable to represent the continuous values in an ordered form.

\noindent \textbf{Precision}.


The continuous numerical constraints with different precisions during pre-training have an impact on the model performance based on Eq.\ref{equation_end}. To this end, we design ablation experiments to explore the specific effects. With all other conditions consistent, the precision grids 0.1, 0.01, and 0.001 are created to pre-train the model. The results are shown in Table \ref{tab:ablation}, where the performance of the model at 0.01 exceeds 0.1 and 0.001  by a greater margin. The reason for such a phenomenon is mainly that the model needs to balance the precision of accuracy and perturbation. As shown in Table \ref{tab:overall}, the MSE loss of various DRP methods is around 0.005, and practically the model cannot accurately predict to the thousandth part, and there is a probability that the value of the thousandth part will be a perturbation factor during the contrastive learning of the text. Therefore, the different performance of DRP models needs to be selected for pre-training with suitable precision.


\noindent \textbf{Caption}.


In the pre-training phase, the prompt of the caption encoder contains the inputs of the drug response process and the numerical results ($IC_{50}$). Since the inputs contain drugs and cell lines, to verify whether the inputs containing the caption of the drug reaction process improve the ability of the fusion encoder to discretely map different kinds of drug reaction pairs, we removed the text of the inputs to compare the results. As shown in Table \ref{tab:ablation}, the model improves on all metrics with an average improvement of 9.1\% after adding the inputs to the text description. This is due to the fact that the description of the drug reaction, a real biological phenomenon, requires at least three kinds of information, drug type (drug molecular structure), cell type, and reaction result. Otherwise, the drug reaction process will collapse into a one-dimensional space, which will affect the transformation of the feature space from the caption encoder to the fusion encoder.

\subsection{Algorithm complexity}


When investigating DRP models, the algorithm complexity plays a crucial role, exhibiting pronounced variations among them. For general methods that do not adaptively adjust during the prediction phase, the importance of space complexity is significantly diminished compared to time complexity. This is because such models require training only once but need to be practically applied to predict drug responses for unknown compounds. Faced with billions of unlabeled compounds, the prediction time and accuracy become critical considerations, while the training time and cost of the model can be negligibly small.


\begin{table}[ht!]
\centering
\caption{Algorithmic complexity analysis and performance comparison of different models under the \MN~framework. \textbf{Training Time} signifies the duration required for processing one batch, while \textbf{Gain} denotes the magnitude of performance enhancement within the \MN.}
\label{tab:alg_comlexity}
\renewcommand \arraystretch{1.3}
\setlength{\tabcolsep}{2mm}{
\resizebox{\columnwidth}{!}{%
\begin{tabular}{ccccc}
\hline
Model & Complexity & \begin{tabular}[c]{@{}c@{}}Training Time\\ (s)\end{tabular} & \begin{tabular}[c]{@{}c@{}}Params\\ (M)\end{tabular} & \begin{tabular}[c]{@{}c@{}}Gain\\(\%)\end{tabular} \\ \hline
tCNNs & $\mathcal{O}(9knd^2)$ & 2.35 & 233 & 7.88 \\
DeepCDR & $\mathcal{O}(7knd^{2}+4n^{2})$ & 1.69 & 179 & 17.83 \\
DeepTTC & $\mathcal{O}(8knd^{2}+dn^{2})$ & 1.28 & 1573 & 19.49 \\
GraphDRP & $\mathcal{O}(3knd^{2}+(2+d)n^{2}+3n)$ & 1.58 & 529 & 14.29 \\
GratransDRP & $\mathcal{O}(3knd^{2}+(2+3d)n^{2}+3n)$ & 1.71 & 2361 & 13.04 \\
TransEDRP & $\mathcal{O}((5d+2)n^{2}+3n)$ & 0.98 & 2244 & 31.46 \\ \hline
+CLDR & $+\mathcal{O}(2n^{2}+3n)$ & +4.21 & +5811 & +17.3 \\\hline
\end{tabular}%
}
}
\end{table}

The experiment utilized an Intel Xeon E5-2690 v3 processor with 12 cores (24 threads) and a clock frequency of 2.60GHz. Additionally, the RTX 4090 GPU is employed. In the \textbf{Training Time} experiments, each model undergoes testing 10 times in the GPU environment. As shown in Table. \ref{tab:alg_comlexity}, the time complexity introduced by the \MN~framework shows a relatively minor variation, $+\mathcal{O}(2n^{2}+3n)$. The TransEDRP model stands out with a remarkable training time per batch of 0.98 seconds, showcasing superior efficiency compared to other models. In contrast, the tCNNs model requires a relatively extended training time, clocking in at 2.35 seconds. The incorporation of the \MN~significantly extends both training time and parameter count, by an additional 4.21 seconds and 5811M, respectively. Notably, the \MN~demands fewer training epochs in both the pre-training and fine-tuning phases, resulting in an overall shorter training duration compared to the original models. In summary, the integration of the \MN~, while contributing to an increased parameter count, substantially enhances generalization performance. Our work represents a breakthrough when compared to traditional DRP methods.

\subsection{Case Study Analysis}

\begin{figure}[ht!]
    \centering
    \includegraphics[width=0.8\linewidth]{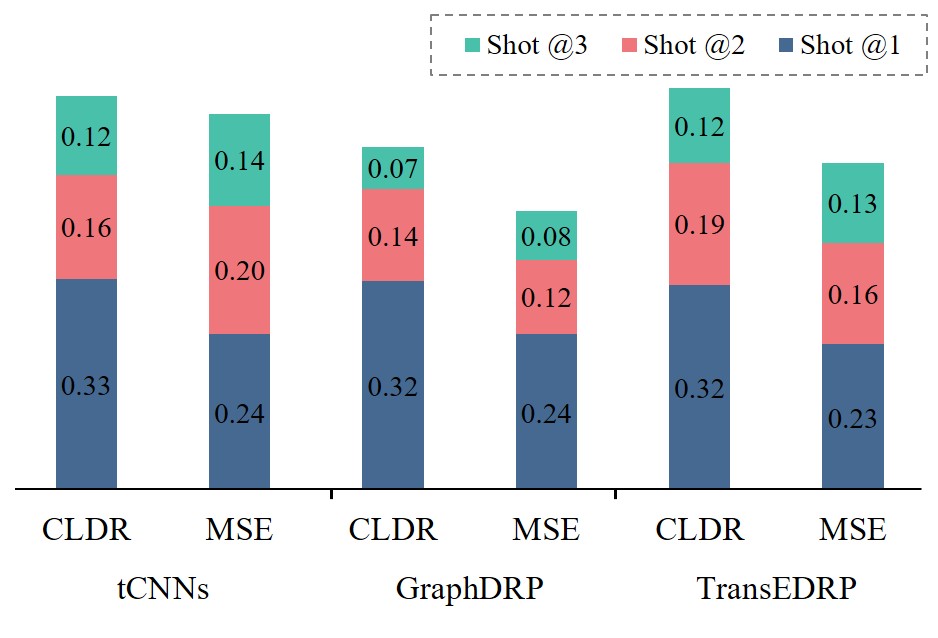}
    \caption{The DRP model's capability in drug recommendation for the identification of cell lines.}
    \label{fig:case}
\end{figure}

In the overall experiments, the \MN~model demonstrates robust accuracy and generalization capabilities, presenting an advanced DRP framework in drug discovery research. To validate that the framework not only exhibits improvements in numerical performance but also genuinely enhances the hit rate in drug screening, we conducted a case study analysis as illustrated in Fig.\ref{fig:case}. We define \textbf{Shot@x} as the probability of hitting the first optimal drug when recommending \textbf{x} compounds. 

\begin{equation}
  Shot@\mathrm{x} = \frac{\sum_{c=1}^{N_{cell}} Hit(\mathrm{T}_{c}^{top_1}, \mathrm{P}_{c}^{top_{\mathrm{x}}})}{N_{cell}} 
\end{equation}
\noindent where $\mathrm{T} and \mathrm{P}$ denote the compounds with true labels and predicted values ranked 1/$\mathrm{x}$ for the $c$ cell line, respectively, and the $Hit(\cdot)$ is 1 when the inputs are equal and 0 otherwise.

We present the capabilities of three representative methods, tCNNs, GraphDRP, and Trans EDRP, both under the original strategy (MSE) and using the CLDR framework, in real-world scenarios for recommending effective compounds tailored to a specific cell line.

We compared the predicted $IC_{50}$ rankings of multiple drugs by the model with the order of the actual values and calculated the probability of hitting the first optimal drug when recommending $\mathrm{x}$ compounds. With the application of the \MN framework across 984 cell lines in the GDSC2, the probability of the model hitting the first optimal drug is consistently above 32\%. In other words, for a given set of 23 compounds, our model achieves a one-shot accuracy of approximately 32\% across around 320 cell lines, and by the second attempt, it accumulates a probability between 46\% and 51\%. 


\section{Conclusion}

In this paper, we propose the \MN, a contrastive learning framework with natural language supervision for the DRP. The \MN~converts regression labels into text, which is merged with the captions text of the drug response process as a second modality for the samples in addition to the traditional encoding modality. In order to enhance the continuous representation capability of the numerical text, the CN-KG is proposed to constrain the ability of the caption encoder to perceive continuous values. We provide rigorous theoretical proof of the soundness of the \MN~method and conduct validation experiments on the GDSC2. The \MN~effectively constructs links between drug response data and labeled text, which in turn enhances the out-of-distribution generalization ability and improves the success rate of phenotypic drug discovery.




\newpage
\bibliographystyle{named}
\bibliography{ijcai22}

\end{document}